\documentclass[twocolumn]{aastex631}

\newcommand{\ltsima} {$\; \buildrel < \over \sim \;$}
\newcommand{\gtsima} {$\; \buildrel > \over \sim \;$}
\newcommand{\lta} {\lower.5ex\hbox{\ltsima}}
\newcommand{\kmsMp}{km s$^{-1}$\,Mpc$^{-1}$}
\newcommand{\gta} {\lower.5ex\hbox{\gtsima}}

\newcommand{\kms}{km\,s$^{-1}$}

\newcommand{\cip}{[C\,{\sc i}]~$^{3}$P$_{1}$-$^{3}$P$_{0}$}
\newcommand{\ci}{[C\,{\sc i}]}
\newcommand{\cii}{[C\,{\sc ii}]}

\begin{document}

\title{Massive molecular outflow and 100 kpc extended cold halo gas in the enormous Ly$\alpha$ nebula of QSO 1228+3128}

\correspondingauthor{Zheng Cai}
\email{zcai@mail.tsinghua.edu.cn}

\author{Jianrui Li}
\affil{Department of Astronomy, Tsinghua University, Beijing, China, 100084}

\author{Bjorn\,H.\,C. Emonts}
\affil{National Radio Astronomy Observatory, 520 Edgemont Road, Charlottesville, VA 22903}

\author{Zheng Cai}
\affil{Department of Astronomy, Tsinghua University, Beijing, China, 100084}

\author{J. Xavier Prochaska}
\affil{UCO/Lick Observatory, University of California, 1156 High Street, Santa Cruz, CA 95064, USA}
\affil{Kavli Institute for the Physics and Mathematics of the Universe (WPI), The University of Tokyo, Kashiwa 277-8583, Japan}

\author{Ilsang Yoon}
\affil{National Radio Astronomy Observatory, 520 Edgemont Road, Charlottesville, VA 22903}

\author{Matthew\,D. Lehnert}
\affil{Sorbonne Universit\'{e}, CNRS, UMR 7095, Institut d'Astrophysique de Paris, 98bis bvd Arago, 75014, Paris, France}

\author{Shiwu Zhang}
\affil{Department of Astronomy, Tsinghua University, Beijing, China, 100084}

\author{Yunjing Wu}
\affil{Department of Astronomy, Tsinghua University, Beijing, China, 100084}

\author{Jianan Li}
\affil{Department of Astronomy, Tsinghua University, Beijing, China, 100084}

\author{Mingyu Li}
\affil{Department of Astronomy, Tsinghua University, Beijing, China, 100084}

\author{Mark Lacy}
\affil{National Radio Astronomy Observatory, 520 Edgemont Road, Charlottesville, VA 22903}

\author{Montserrat Villar-Mart\'{i}n}
\affil{Centro de Astrobiología (CSIC/INTA), Instituto Nacional de Técnica Aeroespacial, l Ctra de Torrejón a Ajalvir, km 4, 28850 Torrejón de Ardoz (Madrid), Spain}



\begin{abstract}
The link between the circum-galactic medium (CGM) and the stellar growth of massive galaxies at high-$z$ depends on the properties of the widespread cold molecular gas. As part of the SUPERCOLD-CGM survey (Survey of Protocluster ELANe Revealing CO/\ci\ in the Ly$\alpha$-Detected CGM), we present the radio-loud QSO Q1228+3128 at $z=2.2218$, which is embedded in an enormous Ly$\alpha$ nebula. ALMA+ACA observations of CO(4-3) reveal both a massive molecular outflow, and a more extended molecular gas reservoir across $\sim$100 kpc in the CGM each containing a mass of M$_{\rm H2}$\,$\sim$\,4$-$5\,$\times$\,10$^{10}$ M$_{\odot}$. The outflow and molecular CGM are aligned spatially, along the direction of an inner radio jet. After re-analysis of Ly$\alpha$ data of Q1228+3128 from the Keck Cosmic Web Imager, we found that the velocity of the extended CO agrees with the redshift derived from the Ly$\alpha$ nebula and the bulk velocity of the massive outflow. We propose a scenario where the radio source in Q1228+3128 is driving the molecular outflow and perhaps also enriching or cooling the CGM. In addition, we found that the extended CO emission is nearly perpendicular to the extended Ly$\alpha$ nebula spatially, indicating that the two gas phases are not well mixed, and possibly even represent different phenomena (e.g., outflow vs. infall). Our results provide crucial evidence in support of predicted baryonic recycling processes that drive the early evolution of massive galaxies.
\end{abstract}


\section{Introduction}\label{intro:}
At high redshifts, large reservoirs of circum-galactic medium (CGM) are known to play a crucial role in the evolution of massive galaxies. Giant gaseous halos, generally detected in Ly$\alpha$ ($T > 10^{4}$K), have been known to be associated with massive galaxies in high-$z$ proto-clusters since the early 2000s \cite[e.g.,][]{Reuland2003, Villar_Martin2003, Miley2006, Cai2017b, Cai2018}. 
Nevertheless, a limitation in our understanding of the multi-phase CGM around distant galaxies is that we have little information about the coldest phase of the CGM. This means that a direct connection to the stellar growth of massive galaxies remains missing, until we identify the ultimate reservoir of halo gas that can fuel widespread star-formation, namely the cold molecular gas ($\sim$10-100 K).

This gap in our understanding started to be filled when large amounts of widespread, cold molecular gas were detected in CO, \ci, and \cii\ on scales of many tens of kpc in the halo environments of massive high-$z$ galaxies. \cite[e.g.,][]{Emonts2014, Emonts2015, Emonts2016, Emonts2018, Cicone2015, Cicone2021, Ginolfi2017, Frayer2018, Fujimoto2019}. The best studied example is the Spiderweb Galaxy, which contains a giant ($\sim$250 kpc) Ly$\alpha$ halo \citep{Miley2006}. Sensitive surface-brightness observations of CO(1-0), CO(4-3) and \cip\ revealed a widespread ($\sim$70 kpc) reservoir of cold molecular gas across the CGM of the Spiderweb \citep{Emonts2016, Emonts2018}. 

This cold CGM has a carbon abundance and excitation properties similar to the ISM in star-forming galaxies \citep{Emonts2018}, and follows diffuse blue light from in-situ star formation in the CGM \citep{Hatch2008, Emonts2016}. This suggests that recycling and mixing through metal-enriched outflows or mass transfer among galaxies, as well as star formation, occurs on a massive scale in the CGM of the Spiderweb \citep{Narayanan2015, Faucher_Giguere2016}. Recently, molecular gas on even larger scales was detected in CO(3-2) across the CGM around a quasar at $z$\,$\sim$\,2 \citep{Cicone2021}. If other high-$z$ proto-cluster galaxies show similar nebulae with cold star-forming gas, then this requires us to revise our ideas of star formation, and the early build-up of the most massive galaxies in the Universe. 
      
Driven by the major advances in optical integral-field-unit (IFU) spectrographs, a new sample of giant Ly$\alpha$-emitting nebulae reminiscent of the Spiderweb, and sufficiently luminous for quantitative analysis of diffuse gas emission, was discovered. 
These “Enormous Ly$\alpha$ Nebulae (ELANe)” are the extrema of Ly$\alpha$ nebulae at $z\sim$2-3, with sizes exceeding the diameters of even massive dark matter halos ($\sim$250 kpc) and Ly$\alpha$ luminosities $>$10$^{44}$ erg/s \citep{Cai2017a, Cantalupo2014, Arrigoni_Battaia2019, Hennawi2015, Borisova2016}. They generally contain one or more luminous AGN, and are typically located within proto-clusters. To trace the molecular CGM across ELANe, the Karl G. Jansky Very Large Array (VLA) was used to observe the enigmatic Mammoth ELAN \citep{Emonts2019}. These data reveal
a large reservoir of cold molecular gas ($\sim$6 $\times$10$^{10}$ M$_{\odot}$) that is spread across tens of kpc in the CGM, even in the presence of an obscured Quasi-Stellar Object (QSO). This suggests that a molecular CGM may occur among the general population of ELANe. However, in other ELANe, such a molecular CGM was not detected \citep{Decarli2021}.

To study the CGM around active galaxies in the Early Universe in a systematic way, we started a survey with the Atacama Large Millimeter/submillimeter Array (ALMA), ALMA Compact Array (ACA), and Karl G. Jansky Very Large Array (VLA) entitled SUPERCOLD-CGM: “Survey of Protocluster ELANe Revealing CO/CI in the Ly$\alpha$ Detected CGM”. The goal of the SUPERCOLD-CGM survey is to use radio telescopes with short baseline configurations to image low-surface-brightness CO and \ci\ emission of widespread cold molecular gas around the massive host galaxies of active galactic nuclei (AGN) at high-$z$. Our initial ALMA/ACA sample consists of 10 type-I quasi-stellar objects (QSOs) at $z\sim$2, which were selected from the SDSS-IV/eBOSS sample. \citep[e.g.,][]{Paris2017} These QSOs contain Enormous Ly$\alpha$ Nebulae (ELANe) mapped by the Keck Cosmic Web Imager (KCWI) \citep{Cai2019}. Here we present observations sensitive to low-surface-brightness emission of molecular gas, CO(4-3), of Q1228+3128 at $z=2.2218$. Q1228+3128 is the most extended and brightest Ly$\alpha$ nebula in our sample of ELANe. 

We assume a ${\Lambda}$CDM cosmology with H$_{0} = 70$\,\kmsMp, $\Omega_\textrm{M} = 0.30$ and $\Omega_{\Lambda} = 0.70$, i.e., 8.25 kpc/$^{\prime\prime}$ and D$_{L}$= 17620 Mpc at $z=2.2218$.

\section{Observation and data reduction}\label{Observations}
We observed Q1228+3128 (12:28:24.97, 31:28:37.70) at $z=2.2218$ for 1.5 hrs on-source during ALMA cycle-7 on 20 and 23 Nov 2019 and 4 Jan 2020 with a compact configuration(C43-2) of the 12m Array (51 antennas), which baselines range from 15 to 314m, as well as for 12.4 hrs between Oct 2019 $\sim$\ Mar 2020 with the ACA 7m Array (12 antennas). We simultaneously centered two adjacent spectral windows of 1.875 GHz on CO(4-3) at $\nu_{\rm obs}$\,$\sim$\,143.10 GHz ($\nu_{\rm rest}$\,=\,461.04 GHz) and another two spectral windows to cover the continuum but also including part of the \cip\ line, which is also detected but much weaker and fell at the outer edge of the band. The \ci\ results will be discussed in a future paper.

The ALMA data were reduced in CASA 5.6.1-8 (Common Astronomy Software Applications; \cite{McMullin2007}). To calibrate the data, we used the archival calibration script supplied by ALMA. After the calibration, we subtracted the continuum from the data by fitting a straight line to the line-free channels in the $uv$-domain, and then combined the respective 12m and 7m datasets in preparation for imaging in tclean\footnote{https://casadocs.readthedocs.io/en/stable/api/tt/casatasks.\ imaging.tclean.html}.

In the step of tclean, we imaged the 12m data using natural weighting, and also imaged the combined 12m+7m
datasets using various degree of tapering\footnote{Tapering means applying a Gaussian taper to the weights of the $uv$-data during imaging, which increases the weighting of the shorter baselines. This decreases spatial resolution and increases the surface brightness sensitivity for detecting extended emission.}. In this paper, we present the combined 12m+7m data tapered to the baseline of 39m. 
We also present the untapered (full resolution) data derived from the continuum-subtracted 12m Array. In addition, we obtained multi-frequency synthesis images of the 150 GHz rest-frame continuum by imaging the line-free channels of the 12m and 7m data using natural weighting, which revealed an unresolved point source with flux density $S_{\rm 150 GHz}$\,=\,8.1\,$\pm$\,0.5 mJy\,beam$^{-1}$. 

For the tapered and untapered CO(4-3) line-data sets, we
imaged our field out to $\sim$70$^{\prime\prime}$ and
binned both of them to 60 \kms\ per channel. We then
convolved the tapered CO(4-3) datacube with a
5.0$^{\prime\prime}\times$5.0$^{\prime\prime}$ Gaussian
to obtain a smoothed synthesized beam of
8.0$^{\prime\prime}\times$7.5$^{\prime\prime}$ with
PA=24.2$^{\circ}$. The synthesized beam size of the
untapered CO(4-3) datacube is
2.8$^{\prime\prime}\times$2.0$^{\prime\prime}$ with
PA=13.6$^{\circ}$. The root-mean-square (rms) noise of
the untapered and smoothed tapered CO(4-3) datacubes is
0.17 and 0.32 mJy\,beam$^{-1}$ channel$^{-1}$,
respectively. 
To get position-velocity (P-V) maps, we binned our smoothed tapered CO(4-3) datacube to 180 \kms\ per channel, resulting in a noise of 0.18 mJy\,beam$^{-1}$ channel$^{-1}$.

To check the validity of the features that we describe in this paper, we also performed a 
self calibration using the model of the continuum image. All features that we describe in this paper where still present in the self-calibrated datacubes. However, since the continuum is faint, our data quality did not improve after the self calibration. Therefore we present in this paper the data products without self calibration applied. Also, we found that the flux density of the 12m and 7m data was consistent when comparing the continuum image of the unresolved radio continuum in the untapered 12m data and the smoothed tapered 12m+7m data, which confirmed that the flux calibration between the different data sets is accurate. 

\begin{deluxetable*}{lccccccc}
\tablecaption{Results of the Gaussian fits to the CO(4-3) emission line spectrum of the QSO and extended region \label{tab:gauss_fit}}
\tablehead{\colhead{Component} & \colhead{$v$}  &  \colhead{$\sigma_v$}  &  \colhead{$I_{\nu}$}  &  \colhead{$L^{\prime}_{\rm CO(4-3)}$} & \colhead{M$_\mathrm{H_2}$} \\ 
\colhead{} & \colhead{[\kms]} & \colhead{[\kms]} & \colhead{[Jy \kms]} & \colhead{[10$^{10}$ K \kms\ pc$^{2}$]} & \colhead{10$^{10}$ M$_{\odot}$} }

\startdata
Narrow & -21 $\pm$ 17  & 130 $\pm$ 20 & 0.64 $\pm$ 0.13 &0.94 $\pm$ 0.19&3.4 $\pm$ 0.7\\
Broad & -346 $\pm$ 130&495 $\pm$ 96&0.79 $\pm$ 0.18&1.17 $\pm$ 0.26&4.2$\pm$ 0.9\\
Extended emission &-413 $\pm$ 45&182 $\pm$ 44&0.43 $\pm$ 0.09 &0.62 $\pm$ 0.13&5.0 $\pm$ 1.0
\enddata
\end{deluxetable*}

\section{Results}\label{Results}
\begin{figure*}
\centering
\includegraphics[width=1.0\textwidth]{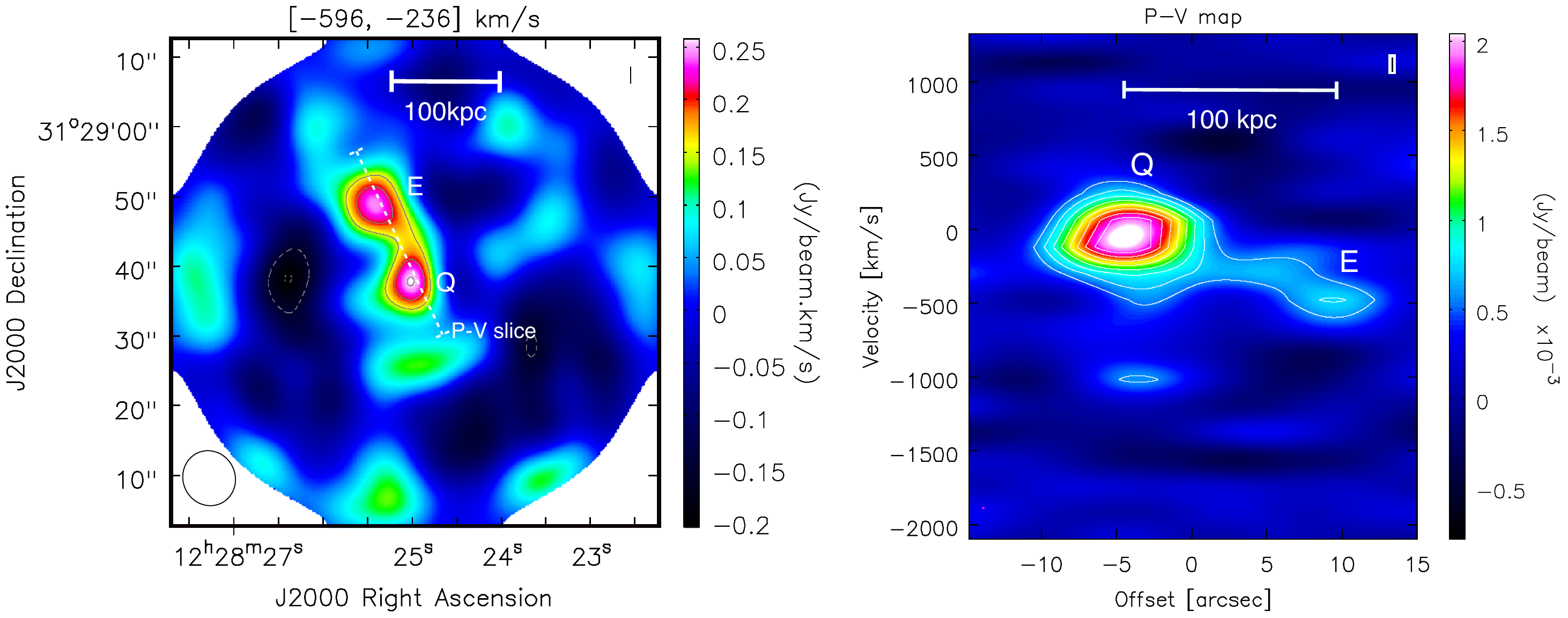}
\caption{{\sl left}: Total intensity map collapsed from the smoothed tapered CO(4-3) datacube integrated within $(-596<v< -236)$ \kms and 1$\sigma$ is 0.05 Jy beam$^{-1}$ km s$^{-1}$. {\sl right}: Position-Velocity (P-V) map extracted along the P-V slice-direction indicated in above panel with 1$\sigma$ equaling 0.18 mJy\,beam$^{-1}$ channel$^{-1}$. In both panels, contours start at 3$\sigma$ and increase with 1$\sigma$.}
\label{fig:spectrum0}
\end{figure*}

The left panel of Fig.~\ref{fig:spectrum0} shows the 
total intensity map of CO(4-3) collapsed from the smoothed tapered datacube integrated within the velocity range of $-596<v< -236$ \kms, and the right panel shows the Position-Velocity (P-V) map extracted along the P-V slice-direction indicated in the left panel. Both of them revealed a very extended reservoir of CO, which is spreading more than 100 kpc in NE direction. "Q" and "E" resemble the QSO and the extended region respectively.

\begin{figure*}
\centering
\includegraphics[width=1.0\textwidth]{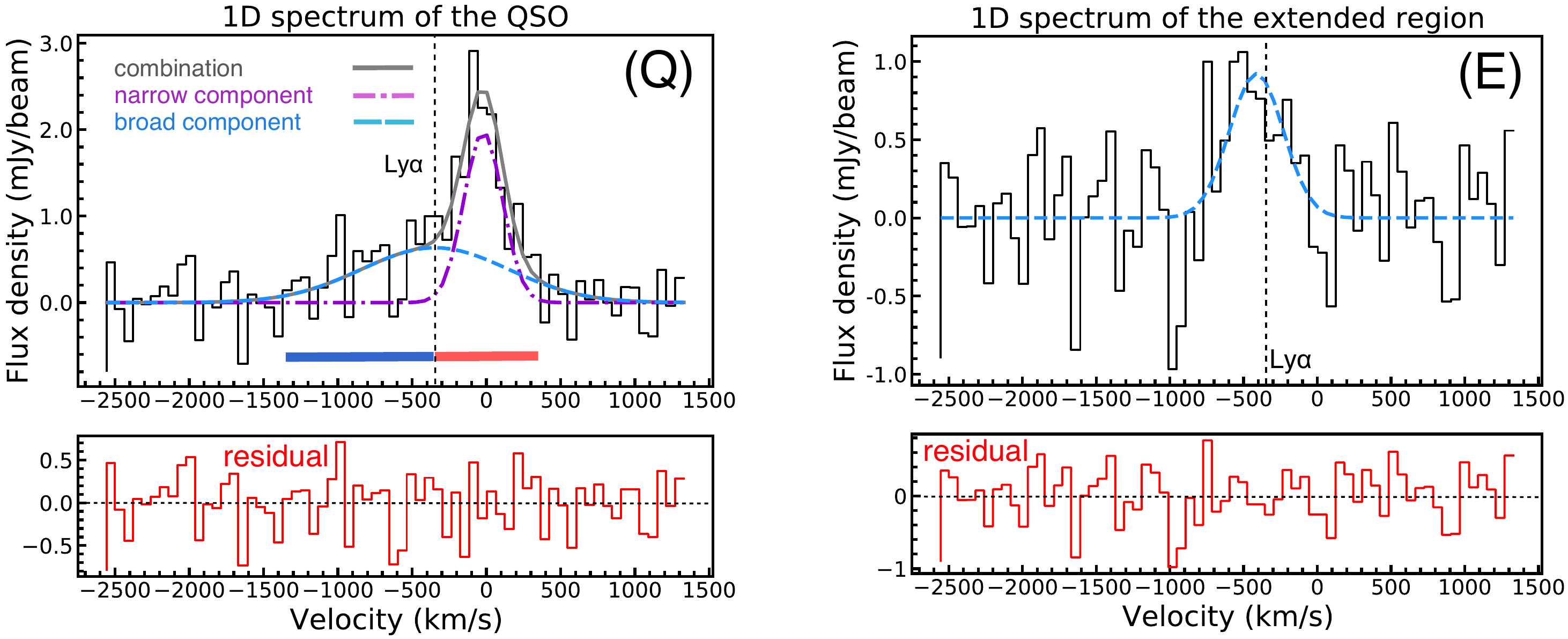}
\caption{{\sl left}: 1D spectrum of the beam-integrated CO(4-3) flux extracted from the smoothed tapered datacube against the peak of the CO emission associated with the quasar. A double Gaussian fit  has been applied to the line profile(black), including a narrow systemic component which traces quiescent gas and a broad blueshifted line component which traces the outflow. The grey line shows the combination of the two Gaussian functions and the red spectrum in the bottom panel shows the residual. The black vertical dash line shows the mean redshift of diffuse Ly$\alpha$ emission in CO(4-3) frame(the same as the right panel). {\sl right}: 1D spectrum of the beam-integrated CO(4-3) flux extracted from the smoothed tapered datacube after primary beam correction against the peak of the CO emission in the extended region. Also, a Gaussian fit (blue dash line) has been applied to the line profile (black), and the red spectrum shows the residual.}

\label{fig:spectrum1}
\end{figure*}

The left panel of Fig.~\ref{fig:spectrum1} shows the 1D spectrum extracted from the smoothed tapered CO(4-3) datacube against the peak of the CO emission associated with the quasar.

To measure the quasar centroid, we collapsed a total intensity broad-band map integrated within the 
velocity range of $-1217$ \kms\ $<v< 883$ \kms\ from the untapered datacube and then regard the
brightest pixel in CO as the QSO centroid. This is also consistent with the optical coordinate of the
QSO. The spectrum was fitted by a double Gaussian, a narrow component to fit the central core, tracing
the quiescent gas in the QSO, and a broad component to fit the broad wings. The curve fit Python
package was used to do this fitting \footnote{https://docs.scipy.org/doc/scipy/reference/generated/\\scipy.optimize.curve\_fit.html}. We chose the area, center and sigma in the Gaussian function as free
parameters, and regarded the 1 / rms of each channel as the weighting. The broad component in
Fig.\,\ref{fig:spectrum1} (left) has a full width at half the maximum intensity of FWHM = 1160
km\,s$^{-1}$, which is significantly larger than what is typically for high-$z$ QSOs and Sub-Millimeter
Galaxies \citep{Carilli_Walter2013}. We therefore argue that this broad component is most likely a
massive outflow of molecular gas.

The right panel of Fig.~\ref{fig:spectrum1} shows the 1D spectrum extracted against the peak of the CO(4-3) emission in the extended region from the smoothed tapered data cube after primary beam correction. The spectrum was fitted by a Gaussian Function (blue dash line) in the same way described above. 
The fitting results and the corresponding luminosity and gas mass are reported in Table\,\ref{tab:gauss_fit}. Fig.~\ref{fig:uv} in Appendix.2 shows an analysis of the extended CO emission in the $uv$-domain by plotting the $uv$-distance of the baselines against the real part of the visibility amplitudes in the extended region \citep{Ivison2010, Cicone2021}. With the length of baselines decreasing, the flux increases conspicuously. From the Gaussian fitting(red dash line in Fig.~\ref{fig:uv}), we derived a FWHM = (8.8 $\pm$ 1.7) $^{\prime\prime}$ of the extended CO emission in the image plane, which corresponds to a minimum spatial extent of 73\,$\pm$14 kpc. The zero-baseline flux is 1.03 $\pm$ 0.25 mJy, which is consistent with the right panel of Fig.~\ref{fig:spectrum1}.

To estimate molecular gas masses, for the QSO, as thermal excitation is reasonable, we assume r$_{\rm 4-3/1-0}$ = $L^{\prime}_{\rm CO(4-3)}$/$L^{\prime}_{\rm CO(1-0)}=1$ \citep{Riechers2011} and for the extended emission we assume r$_{\rm 4-3/1-0}$ = $L^{\prime}_{\rm CO(4-3)}$/$L^{\prime}_{\rm CO(1-0)}=0.45$ \citep{Emonts2018}. The molecular gas mass M$_\mathrm{H_2}$ is derived from $L'_{\rm CO(1-0)}$ \citep{Solomon_Vanden_Bout2005}, assuming $\alpha_{\rm CO}$\,=\,3.6 M$_{\odot}$ (K \kms\ pc$^{2}$)$^{-1}$ \citep{Bolatto2013, Daddi2010, Genzel2010}. In this case, the molecular gas mass M$_\mathrm{H_2}$of the QSO host galaxy, outflow and extended region that we derived from the smoothed tapered datacube is (3.4 $\pm$ 0.7), (4.2 $\pm$ 0.9) and (5.0 $\pm$ 1.0) $\times$ 10$^{10}$ M$_{\odot}$, respectively.

\begin{figure*}
\centering
\includegraphics[width=1.0\textwidth]{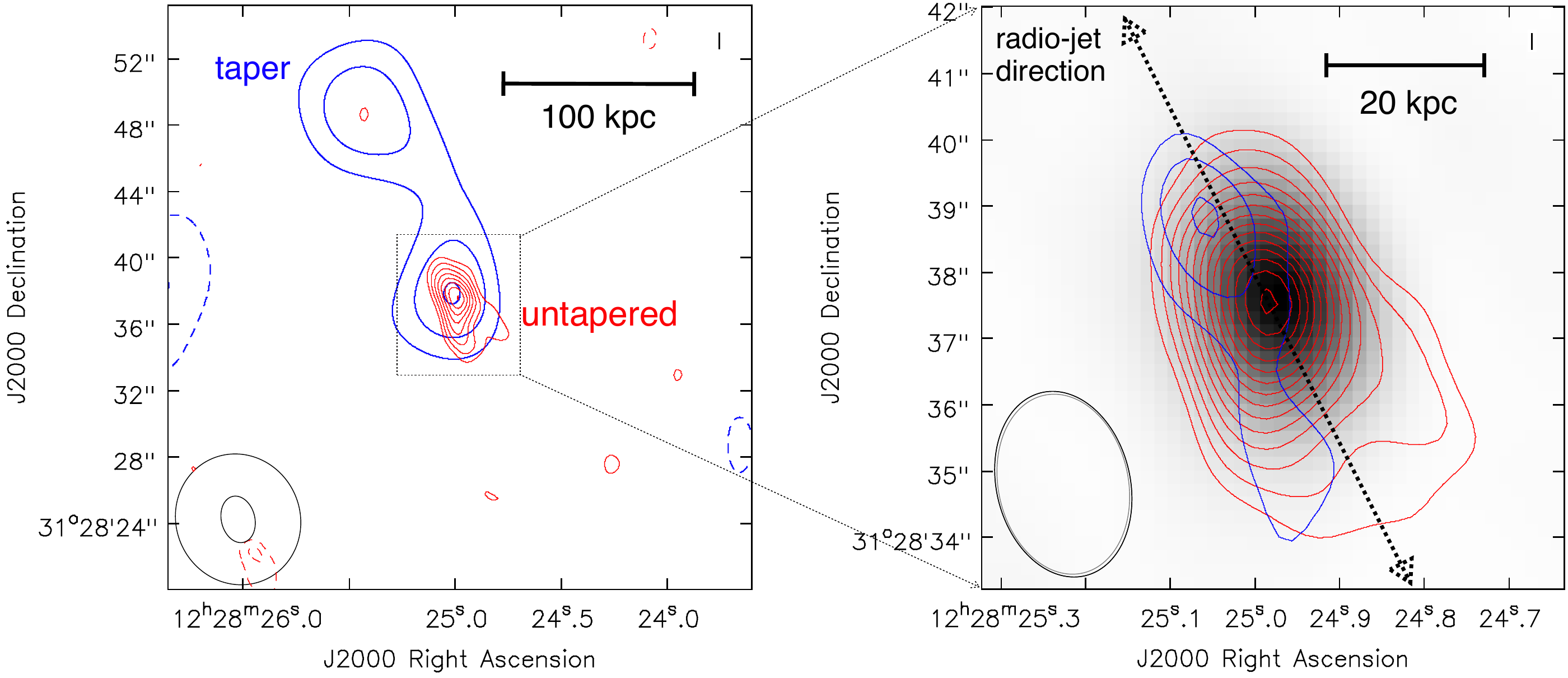}
\caption{{\sl Left:} The blue contours show the total intensity CO(4-3) map from Fig. \ref{fig:spectrum0}. The red contours show the total intensity map collapsed from the untapered CO(4-3) datacube integrated within $v\in(-1217, 883)$ \kms and 1$\sigma$ is 0.097 Jy beam$^{-1}$ km s$^{-1}$. The synthesized beams of the tapered and untapered data are shown in the bottom-left corner of this panel. {\sl Right:} Blue and red contours: Total intensity contour map collapsed from the untapered CO(4-3) datacube integrated within $v\in(-1217, -377)$ \kms and $v\in(-377, 343)$ \kms with 1$\sigma$ equals 0.046 and 0.047 Jy beam$^{-1}$ km s$^{-1}$ respectively. The velocity ranges across which we integrated the systemic and outflow signal are indicated with a red and blue bar in the quasar spectrum in Fig.~\ref{fig:spectrum1}. The background shows the ALMA continuum image. In both panels, contours start at 3$\sigma$ and increase with 1$\sigma$. The black dashed arrow represents the radio-jet direction\citep{Helmboldt2007}.}
\label{fig:spectrum2}
\end{figure*}

The left panel of Fig.~\ref{fig:spectrum2} shows the extended emission at two different resolutions. The red contours represent the total intensity map collapsed from all the line emission channels of the full-resolution (untapered) CO(4-3) datacube, which is spread across $\sim$25 kpc. The blue contours represent the total intensity map collapsed from the blue side of $-596<v< -236$ \kms\ of the smoothed tapered datacube, revealing emission on scales of more than 100 kpc. 
The right panel of Fig.~\ref{fig:spectrum2} shows the two components of the untapered CO(4-3) datacube separated spatially. The red and blue contours represent the quiescent gas in the QSO host galaxy and the blue-sided outflow, respectively. There is an offset of $\sim$15 kpc between the central QSO component and the peak of the CO emission in the outflow. We can make a rough estimate of the mass flow rate of the outflow if assume that the outflow originated from the QSO core and traveled at a velocity of $\Delta v$\,+\,0.5\,$\times$\,(FWHM$_{\rm outflow}$), with $\Delta v$ the velocity difference between the peak velocity of the narrow and broad component and FWHM\,=\,2.35$\sigma_{v}$ (Table \ref{tab:gauss_fit}). This results in an estimated outflow rate of \.{M}$_{\rm H2}$\,$\sim$\,2600 M$_{\odot}$\,yr$^{-1}$. The background in Fig.\,\ref{fig:spectrum2} shows the ALMA continuum image and the black dash arrow represents the radio-jet direction\citep{Helmboldt2007} which is the same as the outflow direction.

\begin{figure*}
\centering
\includegraphics[width=1.0\textwidth]{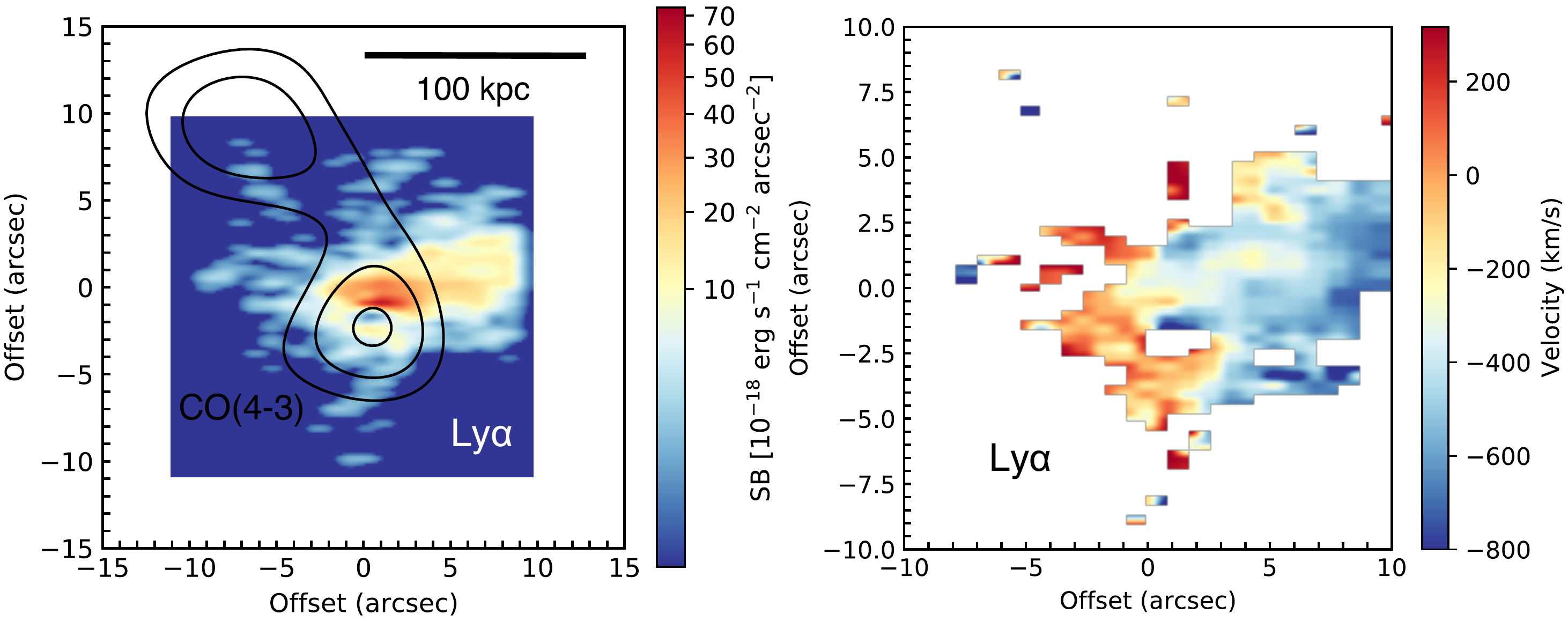}
\caption{{\sl Left:} The black contours show the total intensity CO(4-3) map from Fig.~\ref{fig:spectrum0}, and the background shows the "Optimally extracted" Ly$\alpha$ nebula image from PSF- and continuum-subtracted KCWI data cubes for Q1228+3128 starting at 2$\sigma$, which corresponds to 1.9 $\times$ 10$^{-18}$ erg s$^{-1}$ cm$^{-2}$ arcsec$^{-2}$. {\sl Right:} The kinematics image of the Ly$\alpha$ nebula shown on the left side, with "0 km/s" corresponding to $z=2.2218$.}
\label{fig:spectrum3}
\end{figure*}

The left panel of Fig.~\ref{fig:spectrum3} shows an optimally re-extracted image from PSF- and continuum-subtracted data previously obtained with the Keck Cosmic Web Imager of the Ly$\alpha$ nebula around Q1228+3128 \citep[see][]{Cai2019}. And the right panel of Fig.~\ref{fig:spectrum3} shows the kinematics image of the Ly$\alpha$ nebula in our CO(4-3) frame, with "0 km/s" corresponding to $z=2.2218$, determined from the narrow component of CO(4-3).

\section{discussion}
We have presented large-scale extended CO(4-3) emission spread across $\sim$100 kpc and a massive bluesided outflow from QSO 1228+3128 at $z=2.2218$ as part of the SUPERCOLD-CGM survey. For the central outflow, there is an offset of $\sim15$ kpc between the peak of the systemic CO emission in the host galaxy and the CO in the outflow component, with the outflow found towards the NE (see the right panel of Fig.3). The much more extended and quiescent CO reservoir, found after tapering and smoothing our data, stretches across $\sim100$ kpc in the same direction as the central outflow, and has roughly the same velocity as the bulk velocity of the outflow(see Fig.2). This demonstrates that the molecular outflow and the extended, quiescent molecular gas reservoir may have the same origin. After re-analysis of Ly$\alpha$ data of Q1228+3128 from the Keck Cosmic Web Imager, we converted the Ly$\alpha$ emission into our CO(4-3) redshift frame and found that the mean redshift of Ly$\alpha$ nebula is close to -350 km/s, which is very consistent with the bulk velocity of the outflow and extended CO reservoir.

Our case of Q1228+3128 resembles the detection by \citet{Cicone2015} of a massive outflow of \cii\ and a very extended \cii\ reservoir from the
radio-quiet QSO host galaxy SDSS J1148+5251 at $z > 6$, with filamentary structures that have a complex morphology and reach a maximum projected radius of $\sim$ 30 kpc. They found quasar feedback is likely the dominant mechanism driving the outflow, and their observations are qualitatively consistent with radiation-pressure driven dusty shells. Compared to SDSS J1148+5251, the molecular outflow and quiescent CO reservoir appear to be much more collimated in Q1228+3128.

We know from existing radio data from FIRST (Faint Images of the Radio Sky at Twenty-Centimeters) \citep{Becker1994} and VLASS (VLA Sky Survey) \citep{Lacy2020} that Q1228+3128 is the only radio-detected QSO, or quasar, in our 10 targets sample. In addition, as a flat-spectrum radio source, it was shown by \citet{Helmboldt2007} that there is a radio-jet at a scale of a few tens of milli-arcsec in the same direction as our extended CO reservoir and massive outflow(see Fig.~\ref{fig:spectrum2}.)
In high-z radio galaxies (HzRGs), the CO-emission is often aligned with the radio-jets, and found tens of kpc outside the host galaxy \citep{Klamer2004, Emonts2014, Falkendal2021}. 
The most intriguing part of our work is that Q1228+3128 is the first radio source at high-$z$ where both phenomena have been observed simultaneously. 

For our target, the outflow and extended CO reservoir have roughly the same bulk velocity (see Table\,\ref{tab:gauss_fit}), and are both aligned in the same direction. We propose the scenario that the extended CO-emitting gas is formed when the propagating radio-jet drives the outflow, shocks and cools pre-existing dusty halo gas \citep{Gullberg2016, Emonts2016}, and possibly also enriches the gas by dragging metals out into the environment \citep{Kirkpatrick2011}. If this scenario is viable, we can make the testable prediction that a faint extended radio-jet should be found to be aligned with the extended CO(4-3) emission with sensitive surface-brightness (low resolution but high sensitivity on short baselines) radio imaging. 
 
Besides, the diffuse Ly$\alpha$ nebula also owns the same mean velocity as the extended CO reservoir. However, because the spatial distribution of the extended CO and Ly$\alpha$ emission appears to be almost perpendicular, the respective cold and warm gas phases may not represent a well-mixed CGM. \citet{Villar_Martin2003} resolved kinematically the emission from ambient non-shocked gas (the quiescent haloes) and the emission from (jet induced) shocked gas of a sample of 10 high red-shift radio galaxies at $z\sim 2.5$. They argue that the quasar continuum is the dominant excitation mechanism of the quiescent haloes along the radio axis. In our case, if the radio jet is responsible for the cooling and enrichment of gas in the CGM, then photo-ionization of the quasar would be expected to occur roughly along the same direction, which would not explain the perpendicular orientation of the extended CO and Ly$\alpha$ emission. Instead, an alternative scenario is that the Ly$\alpha$ emission may be infalling gas behind the quasar. It is widely recognized that the collisions with cold gas streams could excite Ly$\alpha$ emission \citep{Dijkstra_Loeb2009, Goerdt2010, Rosdahl_Blaizot2012, Vernet2017}. Observationally, \citet{Daddi2021} performed imaging of the Ly$\alpha$ kinematics across a 300 kpc-wide giant Ly$\alpha$ nebula centered on the massive galaxy group RO-1001, which contains three Ly$\alpha$-emitting filaments and may indicate the case for gas infall. An inflow of gas was also proposed by \citet{Vernet2017} to describe the Ly$\alpha$ properties of MRC\,0316-257 at $z$\,$\sim$\,3. The Ly$\alpha$ nebula of Q1228+3128 could present a similar case.

Another alternative scenario is that there is a separate merging galaxy in the extended region, and that the extended molecular gas is tidal debris from galaxy interactions. Although we see a 3$\sigma$ detection of the extended region in the untapered CO(4-3) image, the total flux of the extended region in the untapered CO(4-3) image is only $\sim$40$\%$ of that in the smoothed tapered image, meaning that the bulk of the extended emission is spread on scales that are over-resolved even with the compact 12m ALMA array (see Fig. \ref{fig:uv}). If there is a companion galaxy associated with part of the CO(4-3) emission, it is associated with a non-detection in the ALMA continuum image. When converting the 3$\sigma$ flux density of 0.15 mJy from the untapered continuum image to the IR luminosity($L_{\rm IR\_SF}$) integrated within the wavelength range 8$\sim$1000 $\mu$m, we derived an upper limit for the $L_{\rm IR}$ which is 3.0 $\times$ 10$^{12}$ L$_{\odot}$.\citep{Wang2008} According to the correlation between $L^{\prime}_{\rm CO(4-3)}$ and $L_{\rm IR}$ described in \cite{Greve2014}, we derived a $L_{\rm IR}$ of the extended region which is (4.5 $\pm$ 1.0) $\times$ 10$^{11}$ L$_{\odot}$ from untapered $L^{\prime}_{\rm CO(4-3)}$. The fact that the $L_{\rm IR}$ limit for detecting the continuum is much higher than that calculated for CO(4-3) means we do not have the sensitivity at this frequency to detect the continuum. However, a deep optical image of Q1228 obtained with Keck(see Appendix.1) helped us to investigate the existence of the companion galaxy. Although the QSO was saturated, there is still a non-detection in the extended region. Using the 3$\sigma$ flux density of 3.0 $\times$ 10$^{-30}$ erg s$^{-1}$ cm$^{-2}$ Hz$^{-1}$(g$_{AB}$ $\sim$ 25 mags) from an aperture with diameter equating 2$^{\prime\prime}$, we derived an upper limit of the star formation rate(SFR) for the extended region from the deep optical image which is 2.4 M$_{\odot}$ yr$^{-1}$ \citep{Kroupa2001}. But the $L^{\prime}_{\rm CO(4-3)}$ predicts a SFR of 49 $\pm$ 12 M$_{\odot}$ yr$^{-1}$ \citep{Kennicutt1998}, which is much higher than the optical result. This, combined with the fact that the total amount of molecular gas that must be spread on large scales as part of the tidal debris would be comparatively large ($\sim$3\,$\times$\,10$^{10}$ M$_{\odot}$), suggests that the extended emission is not likely the results of a tidal interaction with a companion galaxy.

\section{Acknowledgement}

Z.C. and J.L. are supported by the National Key R\&D Program of China (grant No. 2018YFA0404503), the National Science Foundation of China (grant No. 12073014) and Tsinghua University Initiative Scientific Research Program (grant No. 2019Z07L02017). MVM work was funded by grant PGC2018-094671-B-I00 funded by MCIN/AEI/10.13039/501100011033 and by the European Union NextGenerationEU/PRTR. We thank the anonymous referee for the very valuable feedback. This paper makes use of the following ALMA data: ADS/JAO.ALMA$\#$2019.1.01251.S. ALMA is a partnership of ESO (representing its member states), NSF (USA) and NINS (Japan), together with NRC (Canada), MOST and ASIAA (Taiwan), and KASI (Republic of Korea), in cooperation with the Republic of Chile. The Joint ALMA Observatory is operated by ESO, AUI/NRAO and NAOJ. The National Radio Astronomy Observatory is a facility of the National Science Foundation operated under cooperative agreement by Associated Universities, Inc.

\appendix

\section{Appendix.1: Deep optical image of Q1228}

\begin{figure}
\centering
\includegraphics[width=0.5\textwidth]{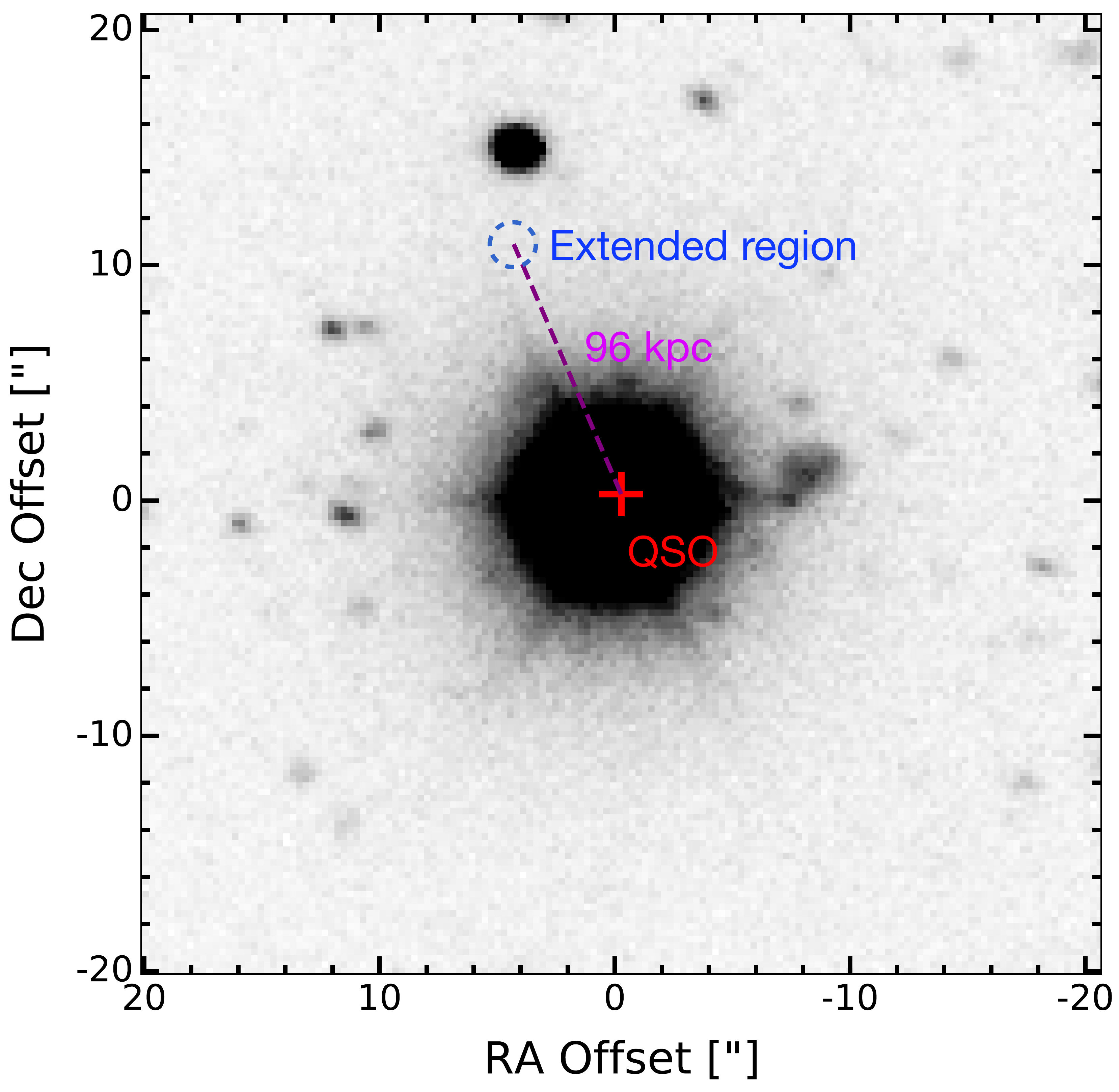}
\caption{The deep optical image of Q1228 from the Low Resolution Imaging Spectrometer (LRIS) of Keck. The blue side G filter covering 416.173 $\sim$ 524.362 nm was chosen for imaging. The small blue dash circle shows the photometry aperture of the extended region, and the red cross shows the QSO.} 
\label{fig:optical}
\end{figure}

Fig.~\ref{fig:optical} shows the deep optical image of Q1228 from the Low Resolution Imaging Spectrometer (LRIS) of Keck. The observation was done on 8 July 2021 with a total on-source time of 25 minutes. The blue side G filter covering 416.173 $\sim$ 524.362 nm was chosen for imaging. We firstly reduced the data in lpipe and then used the SEXTRACTOR, SCAMP and SWARP to do the calibration of WSC and the final stacking process.

\section{Appendix.2: A radial uv plot of the extended region}

\begin{figure}
\centering
\includegraphics[width=0.6\textwidth]{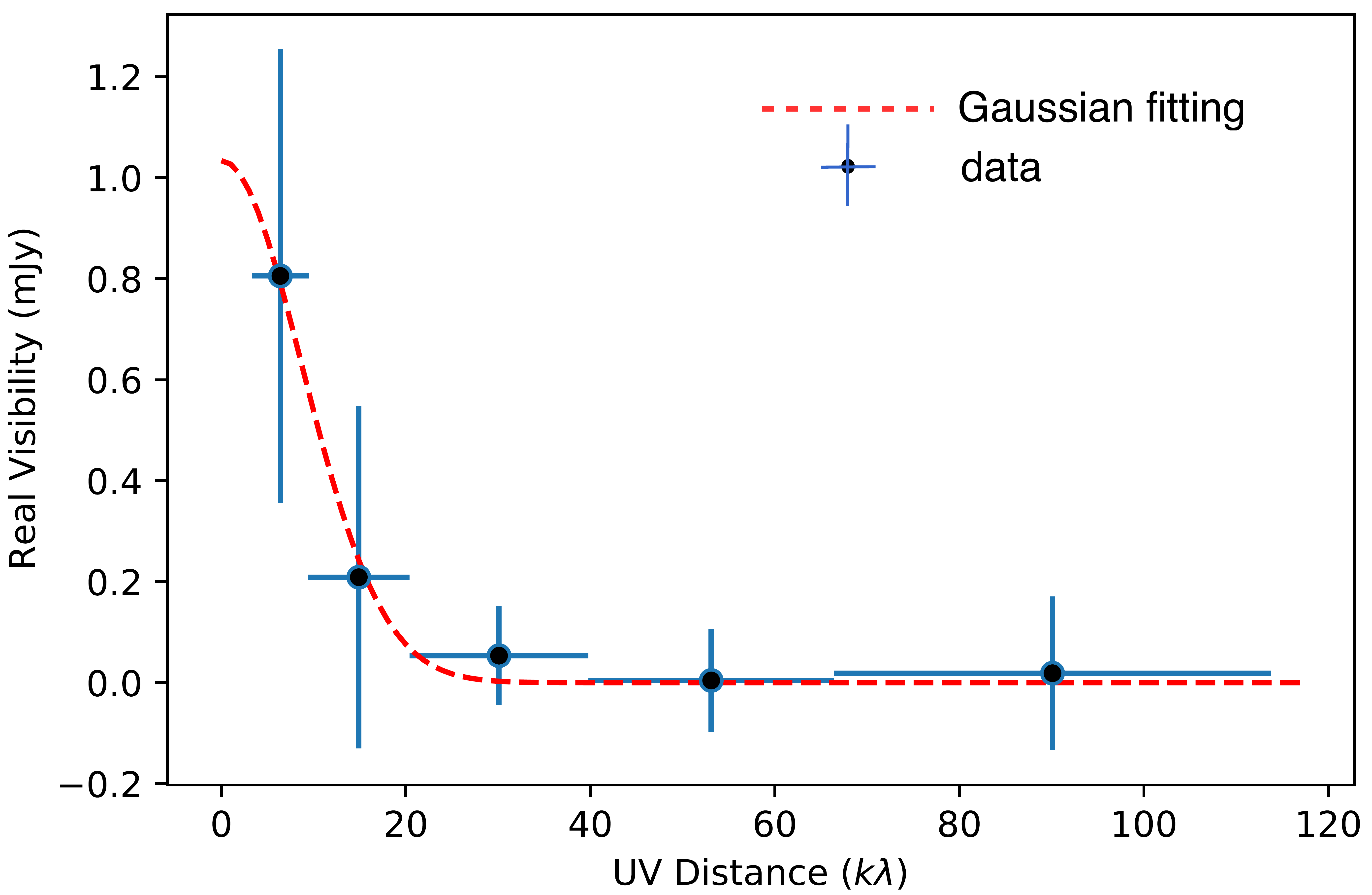}
\caption{$uv$ plot of the 12m and 7m ACA combined CO(4-3) data in the extended region. The vertical and horizontal blue bars represent the error of the flux and the $uv$ distance range per bin respectively. The red dash line shows the fit of a model of a Gaussian distribution with a flux = 1.03 $\pm$ 0.25 mJy and a FWHM = (8.8 $\pm$ 1.7) $^{\prime\prime}$.}
\label{fig:uv}
\end{figure}

Fig.~\ref{fig:uv} shows the $uv$ distance plotted against the real part of the visibility amplitude of the CO(4-3) emission in the extended region. In order to obtain this plot, we first split out the combined 12m+7m datasets to only contain the channel range(143.17$\sim$143.40 GHz) corresponding to the extended emission, and then shifted the phase center from the QSO to the extended region(ICRS: 12:28:25.427, 31.28.48.975). We then extracted all the data points and binned them. Finally, we fitted the Fourier transform of a Gaussian function to the data points, which corresponds to the model of a Gaussian distribution in the image plane.

\bibliographystyle{aasjournal}



\end{document}